\documentclass[a4paper]{jpconf}
\usepackage{amssymb}
\usepackage{epsfig}
\usepackage{amsthm}
\usepackage{latexsym}
\newcommand{\be}{\begin{equation}}
\newcommand{\ee}{\end{equation}}
\newcommand{\bea}{\begin{eqnarray}}
\newcommand{\eea}{\end{eqnarray}}
\newcommand{\bdm}{\begin{displaymath}}
\newcommand{\edm}{\end{displaymath}}

\begin{document}
\title{\textbf{Inflating branes inside topological defects and periodic structures}}
\author{\textbf{Yves Brihaye}}
\date{}
\address{Faculty of Sciences,
University of Mons,
7000 Mons, Belgium}
\ead{\textbf{yves.brihaye@umh.ac.be}}
\author{\textbf{Betti Hartmann \footnote{Speaker}}}
\date{}
\address{Laboratoire de Math\'ematiques et Physique Th\'eorique,
Universit\'e de Tours, Parc de Grandmont, 37200 Tours, France}
\ead{\textbf{betti.hartmann@lmpt.univ-tours.fr}}
 \vspace{1cm}
\begin{abstract}
We study brane world models which contain local topological defects
in the bulk and a $(3+1)$-dimensional inflating brane. We put the emphasis on
new types of solutions that are periodic in the bulk radial coordinate and thus provide examples
of ``naturally compactified'' brane worlds.  
 \end{abstract}

\section{Introduction}
The basic idea of brane worlds \cite{ruba,dvali,anton,arkani,rs1,rs2,akama} is that our Universe is
represented by a 4-dimensional space-time -the brane- embedded in a
higher dimensional space -the bulk. This interpretation raises
many challenges, the main one being the experimental 
signature revealing the extra dimensions. Natural questions are:
how many extra dimensions are there and, if any, are they 
finite, i.e. compact or infinite.

If the extra dimension are large or even infinite, an appropriate
mechanism has to be implemented in order to confine the fields and
their interactions on the brane. Gravity as a property of space-time itself
is naturally not confined to
the brane but the fact that Newton's law is valid down to scales of 
roughly one millimeter has to be taken into account. A possibility is that the
space-time in the bulk is ``warped'' \cite{rs1,rs2}.

Compact extra dimensions as they e.g. appear in Kaluza-Klein theory \cite{kk}
or string theory \cite{string} are implemented into the theory ``by hand''.
However, one could also think that ``natural compactifications'' are possible.
Indeed, a few years ago in the study of the Einstein-Yang-Mills and Einstein-Yang-Mills-Higgs models
in four space-time dimensions it was found that in the presence of a positive cosmological
constant the space-time has a natural compactification for specific values of the
cosmological constant \cite{bfm}.

In this paper, we investigate if similar results are available in the framework
of brane world models. The three main ingredients for the models
studied here are: (i) we have a non-vanishing bulk cosmological 
constant, (ii) there is a local topological
defect residing in the bulk, the nature of which depends on the 
number of extra dimensions $n$: string-like for $n=2$, monopole for $n=3$ and
(iii) the inflating brane is localised at the origin of the 
topological defect. Models containing these ingredients have been
studied in detail in \cite{brihaye_hartmann, brihaye_delsate}, however,
here we put the emphasis on periodic structures and give further details.

Note that our work is an extension of the work in \cite{shapo}, where
brane worlds containing topological defects were studied. However, the branes
were assumed to be static Minkowski branes.
In \cite{cho}, on the other hand, inflating branes were studied, but in a bulk without
cosmological constant.

The paper is organised as follows:   
in Section 2 we describe the model and the ansatz for the metric.
Some analytic solutions of the vacuum Einstein equations are
presented in Section 3. The models containing matter fields
in the bulk in the form of topological defects are presented in Section 4, where we also present
our numerical results. We give our summary in Section 5.

\section{The model}
In this paper, we consider a $(3+1)$-dimensional brane in an $n$-dimensional
space-time. The bulk contains either no matter fields
at all (vacuum case) or a local topological defect. The action for this model reads:
\be
S = \frac{1}{16\pi G_{n+4}}\int \sqrt{-g}
(R + \Lambda_{n+4} )d^{n+4}x  + S_{bulk}
\ee
where $S_{bulk}$ is the action that describes the specific model
and will be given in explicit form
later in the paper. $G_{n+4}
 = \frac{1}{M_*^{n+2}}$ is the fundamental gravity scale
with $M_*$ the $(n+4)$-dimensional Planck mass. $\Lambda_{n+4}$ denotes the bulk cosmological
constant. 

The general form of the non-factorisable
metric that we consider in this paper is given by:
\be
ds^2 = M(r)^2 ds_4^2 + dr^2 + L(r)^2d\Omega_{n-1}
\ee
where $ds_4^2$ describes the metric $g_{\mu\nu}^{(4)}$ of the  $3$-brane. This 4-dimensional
metric satisfies the 4-dimensional Einstein equations:
\be
G_{\mu\nu}^{(4)}= 8\pi G_{4} g_{\mu\nu}^{(4)} \ \ , \ \ \mu,\nu=0,1,2,3  \ .
\ee

 The transverse space has rotational
invariance, $r=\sqrt{\sum\limits_{k=1}^{n}x_k^2}$ describes the bulk radial coordinate, while $d\Omega_{n-1}$ 
denotes the line element associated with 
the $n-1$ angles $\Theta_i$, $i=1,..,n-1$ of the transverse space.

Throughout this paper, we study inflating branes and thus choose
the following Ansatz for the brane metric: 

  \be
   ds_4^2 = - dt^2 + e^{2 H t} \left( d\tilde{x}_1^2 + d\tilde{x}_2^2 + d\tilde{x}_3^2 \right)
   \ee
with $H=\sqrt{\frac{\Lambda_{phys}}{3}}$, where
$\Lambda_{phys}$ denotes the physical cosmological constant on the brane.
$\tilde{x}_1$, $\tilde{x}_2$, $\tilde{x}_3$ are cartesian coordinates on the brane.
The resulting Einstein equations are given in \cite{brihaye_delsate}.

In this paper, we are interested in solutions that have metric functions
which are periodic in the bulk radial coordinate $r$.
These solutions appear for $\Lambda_{phys} > 0$, so we restrict our discussions
here to inflating branes.

\section{Vacuum solutions}
Here, we discuss the solutions of the resulting Einstein equations without
matter fields in the bulk, i.e. $S_{bulk}=0$.

\subsection{$n=2$}
In the case of a two extra dimensions,  it was noticed in \cite{brihaye_hartmann} that the equations lead
to the proportionality of the metric function $L$ to the derivative of $M$, $L\propto M'$.
The metric functions for the case of an inflating brane in a de Sitter bulk then read:
\begin{equation}
\label{trigonometric}
      M(r) = \sqrt{\frac{80 \pi G_4 \Lambda_{phys}}{3 \Lambda_6}} 
      \sin\left(\sqrt{\frac{\Lambda_6}{10}}r\right)  \ \ , \ \
 L(r) = L_0 
      \cos\left(\sqrt{\frac{\Lambda_6}{10}}r \right)  \ .
\end{equation}
where $L_0$ is a constant.

This solution describes a space-time that can be interpreted as infinitely many copies of a finite
space-time with length $2\pi/\sqrt{\frac{\Lambda_6}{10}}$. The finiteness of the space-time allows for a finite
effective 4-dimensional Planck mass $M_{pl}$ \cite{brihaye_hartmann}.

Since the curvature invariants contain a term of the form $1/M^2$, this vacuum solution is singular at
the origin $r=0$. We will see later in the paper that the presence of matter fields in the
bulk can regularise the space-time.

\subsection{$n=3$}
For $n=3$, the Einstein equations lead to $L\propto M$ with the following periodic solutions:
\begin{equation}
\label{trigonometric2}
      M(r) = \sqrt{\frac{24 \pi G_4 \Lambda_{phys}}{ \Lambda_7}} 
      \sin\left(\sqrt{\frac{\Lambda_7}{15}}r\right)  \ \ , \ \
 L(r) = \sqrt{\frac{\Lambda_7}{75}} \sin\left(\sqrt{\frac{\Lambda_7}{15}}r\right)  \ .
\end{equation}
Again, the space-time is periodic in the bulk radial coordinate and can be seen as infinitely many copies
of the finite interval in $r$ with length $2\pi/\sqrt{\frac{\Lambda_7}{15}}$. Like for $n=2$, the space-time
is singular at the origin.

\section{Solutions with non-trivial matter fields in the bulk}
We have seen that periodic solutions are possible, however they are plagued with
a curvature singularity at the origin. To regularise the space-time, we introduce
matter fields in the form of local topological defects in the bulk.

\subsection{$n=2$}
For $n=2$ extra dimensions, we have to choose a topological defect with appropriate
symmetry. A possibility is a string-like defect that has a winding associated
to the map of spatial infinity $S^1_{\infty}$ of the bulk coordinates $r=\sqrt{x_1^2+x_2^2}$, $\theta$ to the
$S^1$ of the corresponding vacuum manifold of the theory. The action $S_{bulk}$ then reads \cite{no,gstring}:
\begin{equation}
S_{bulk}=\int d^6 x \sqrt{-g_6} \left(-\frac{1}{4} F_{MN}F^{MN}
-\frac{1}{2}D_M\phi D^M\phi^*-\frac{\lambda}{4}(\phi^*\phi-v^2)^2  \right) \ , \ M,N=0,1,..,5
\end{equation}
with the covariant derivative $D_M=\nabla_M-ieA_M$ and the field strength
$F_{MN}=\partial_M
A_N-\partial_N A_M$ of the U(1) gauge potential $A_M$.
$v$ is the vacuum expectation value of the complex valued Higgs field $\phi$ and
$\lambda$ is the self-coupling constant of the Higgs field. $e$ denotes the gauge coupling.

The appropriate Ansatz for the gauge and Higgs field reads \cite{no}:
\begin{equation}
\phi(r, \theta)=v f(r) 
e^{i N\theta} \ , \ \ A_{\theta}(r,\theta)=\frac{1}{e}(N-P(r))
\end{equation}
where $N$ is the vorticity of the string, which throughout this paper we choose
$N=1$.

We introduce the dimensionless coordinate $x=\sqrt{\lambda}v r$
and rescale $L\rightarrow \frac{1}{\sqrt{\lambda}v} L$. 
The set of equations then depends only on the following dimensionless coupling constants:
\begin{equation}
\alpha=\frac{e^2}{\lambda} \ , \ \ \ \gamma^2=8\pi G_6 v^2 \ , \ \ \ 
\Lambda=\frac{\hat\Lambda_6}{\lambda v^2} \ , \ \ \  
\kappa=\frac{8\pi G_4\Lambda_{phys}}{\lambda v^2}
\end{equation}

The resulting set of coupled ordinary differential equations is
given in \cite{brihaye_hartmann}.

\subsubsection{Numerical results}

In order to construct solutions numerically, we have to specify
a set of boundary conditions. We require regularity at the origin $x=0$ which leads to the
boundary conditions:
\begin{equation}
\label{bcx0}
f(0)=0 \ , \ \ \ P(0)=1 \ , \ \ \ M(0)=1 \ , \ \ \ M^{'}|_{x=0}=0 \ , \ \ \
L(0)=0 \ , \ \ \  L^{'}|_{x=0}=1 \ .
\end{equation}
Moreover, the matter field functions should reach their vacuum values asymptotically:
\begin{equation}
f(x = \infty)=1 \ , \ \ \ P(x = \infty)=0 \ .
\end{equation}

\begin{figure}[!htb]
\centering
\leavevmode\epsfxsize=9.0cm
\epsfbox{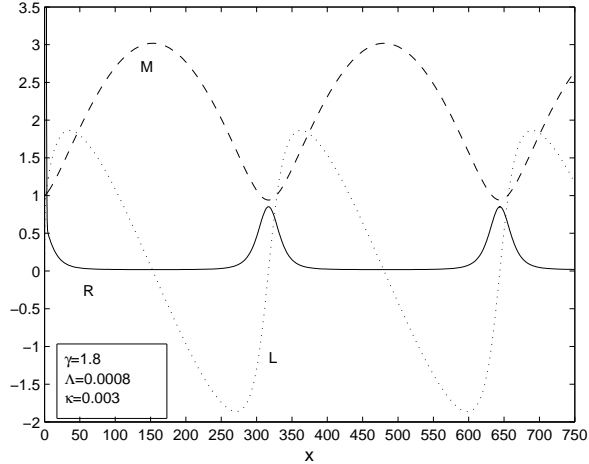}\\
\caption{\label{fig1} 
The metric functions $L$ and $M$ as well as the scalar curvature $R$ are shown as functions of
the radial coordinate $x$ for $\gamma=1.8$, $\Lambda=0.0008$ and $\kappa=0.003$.}
\end{figure}

\begin{figure}[!htb]
\centering
\leavevmode\epsfxsize=9.0cm
\epsfbox{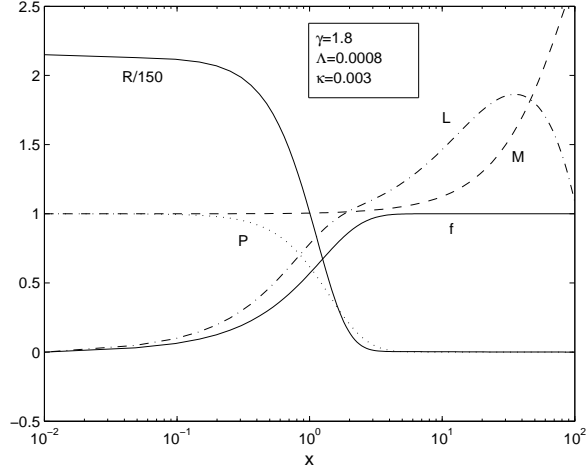}\\
\caption{\label{fig2} 
The metric functions  $L$, $M$, the scalar curvature $R$ as well as the matter field
functions $P$ and $f$ are shown on the interval $x\in [10^{-2}:10^{2}]$ for $\gamma=1.8$, $\Lambda=0.0008$ and $\kappa=0.003$.
Note that the scalar curvature $R$ close to $x=0$ is large, but finite.}
\end{figure}

An example of a periodic solution is given in Fig.s \ref{fig1} and \ref{fig2}.
The presence of the matter fields deforms the periodic solution (\ref{trigonometric}) of the vacuum
case inside the string core. The matter functions $P$ and $f$ reach their vacuum value for
a value of $x$ of order $10$. For larger $x$, the metric functions
behave according to (\ref{trigonometric}). Clearly, the matter fields inside the string core
regularise the periodic solution close to the origin. The Ricci scalar $R$ is large close
to the origin, but stays finite. In such a way, a finite volume brane world that is regular at the origin
appears in this setting. We note that periodic solutions are only possible if the inflating 
branes reside in a de Sitter bulk, i.e. for $\Lambda > 0$.

\subsection{$n=3$}
Here, we choose a local monopole to reside in the bulk. Spatial infinity $S^2_{\infty}$ of the bulk
coordinates $(r, \theta_1, \theta_2)\equiv (r, \theta, \varphi)$ with $r=\sqrt{x_1^2+
x_2^2+x_3^2}$
maps to $S^2$ of the vacuum
manifold of the theory. The action reads:

\be
S_{bulk}= \int  d^7 x \sqrt{-g_7} \left( - \frac{1}{4} F_{MN}^a F^{a,MN} 
-D_M \Phi^a D^M\Phi^a
-\frac{\lambda}{4}(\Phi^a\Phi^a - v^2)^2\right) \ , \ M,N=0,1,..,6 \ .
\ee
with the covariant derivative
$D_M \Phi^a =  \partial_M \Phi^a + e  \epsilon^{abc} A_M^b \Phi^c$ and field strength tensor
$F_{MN}^a = \partial_M A_N^a - \partial_N A_M^a + e  \epsilon^{abc} A_M^b A_N^c$.
and $a=1,2,3$ indexes the components of the real scalar triplet Higgs field $\Phi^a$.
$v$ is the vacuum expectation value of the Higgs field $\Phi$ and
$\lambda$ is the self-coupling constant of the Higgs field. $e$ denotes the gauge coupling.

Along with \cite{cho}, we use a spherically symmetric ``hedgehog'' ansatz for
the gauge and Higgs fields \cite{hooft}:
\be
   A_{\mu}^a = A_r^a= 0 \ \ , \ \ A_{\theta}^a= \frac{1-W(r)}{er} {e}_{\varphi}^a \ \ , \ \
 A_{\varphi}^a= \frac{1-W(r)}{er} \sin\theta {e}_{\theta}^a \  , \ \ a = 1,2,3  \ , 
\ee
\be
    \Phi^a =   \phi(r) e_r^a \ \ \ \ , \ \ a = 1,2,3   \ .
\ee

We introduce the dimensionless coordinate $x=\sqrt{\lambda}v r$
and rescale $L\rightarrow \frac{1}{\sqrt{\lambda}v} L$. The set of coupled differential equations
then depends only on the following dimensionless parameters:
\begin{equation}
\beta=\frac{\lambda}{e^2} \ , \ \ \ \gamma^2=8\pi G_6 v^2 \ , \ \ \ 
\Lambda=\frac{\hat\Lambda_7}{\lambda v^2} \ , \ \ \  
\kappa=\frac{8\pi G_4\Lambda_{phys}}{\lambda v^2}
\end{equation}

The system of coupled equations is given in \cite{brihaye_delsate}

\begin{figure}
\centerline{\psfig{file=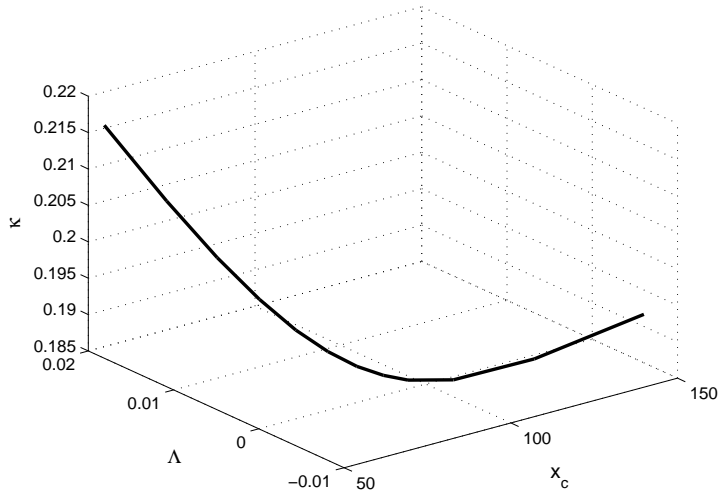,width=10cm}}
\vspace*{8pt}
\caption{\label{fig3} 
The value of $x_c$ for the periodic solutions 
is shown in dependence on the bulk cosmological constant $\Lambda$
and the brane cosmological constant $\kappa$. Note that $\Lambda$ and $\kappa$ have to be
fine-tuned in order to obtain periodic solutions.}
\end{figure}

\subsubsection{Numerical results}
In analogy to the $n=2$ case, one could think
to construct regular counterparts of the vacuum solutions (\ref{trigonometric2}).
However, it turns out that all solutions are singular at some finite value of the
bulk radial coordinate. We have thus
followed a different strategy to construct periodic solutions in $n=3$: we have constructed
solutions on an interval $x\in [0:x_c]$ that can be continued into periodic solutions
on the interval  $x\in [0:2x_c]$. 

We thus impose the following boundary conditions:
\begin{equation}
\phi(0)=0 \ , \ \ \ W(0)=1 \ , \ \ \ M(0)=1 \ , \ \ \ M^{'}|_{x=0}=0 \ , \ \ \
L(0)=0 \ , \ \ \  L^{'}|_{x=0}=1 \ .
\end{equation}
at $x=0$ and 
\begin{equation}
M^{'}|_{x=x_c}=0 \ , \ L^{'}|_{x=x_c}=0   
\end{equation}
at $x=x_c$. $\kappa$ and $\Lambda$ can then be fine-tuned in such a way that
$W(x=x_c)=0$ and $\phi(x=x_c)=0$, i.e. we only find periodic solutions
for specific values of the two cosmological constants. Our results are shown in Fig.\ref{fig3},
where we show the interdependence of $x_c$, $\kappa$ and $\Lambda$ for which periodic solutions
exist. In comparison to the $n=2$ case, periodic solutions are also available for
a negative bulk cosmological constant. Obviously, $x_c$ decreases
with the simultaneous increase of the two cosmological constants.

\begin{figure}

\centerline{\psfig{file=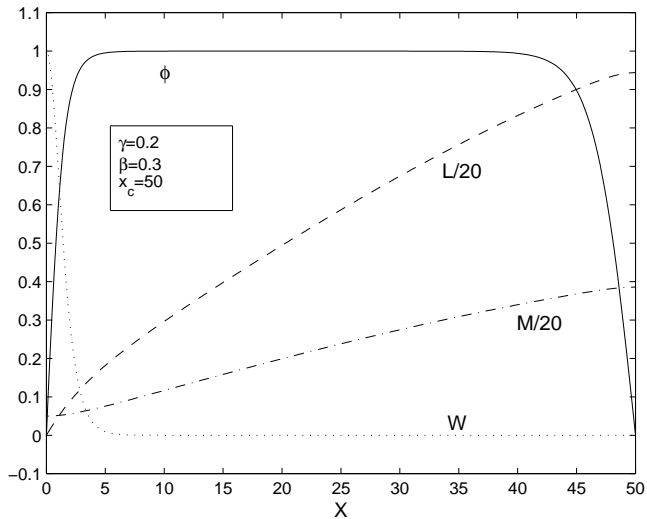,width=10cm}}
\vspace*{8pt}
\caption{\label{fig4} 
The profiles of the metric functions $M$ and $L$ as well as of the matter field functions
$\phi$ and $W$ are shown for a periodic solution with $\gamma=0.2$, $\beta=0.3$
and $x_c=50$. }
\end{figure}

A typical solution for $\gamma=0.2$, $\beta=0.3$ and $x_c=50$ is shown in Fig.\ref{fig4}.
The solutions can then be continued to $x=2x_c$ in such a way that
\bea
M(x_c-x) &=& M(x_c+x) \ , \ L(x_c-x) = L(x_c+x) \ , \ \nonumber\\
\phi(x_c-x) &=& -\phi(x_c+x) \ , \ W(x_c-x) = W(x_c+x)  \ .
\eea

These solutions were coined ``mirror'' symmetric in  \cite{brihaye_delsate}.
Due to the fact that the Higgs field function $\phi$ has a further zero at $x=x_c$, one would expect
a domain wall to lie at this point. Indeed, in  \cite{brihaye_delsate} it was confirmed
that close to $x=x_c$ the equations resemble those of a
kink solution.

Thus, these solutions provide again example of finite brane world models in which
a second ``brane'' naturally appears.


\section{Summary}
In this paper, we have discussed periodic solutions which appear in brane world
model. These solutions appear if an inflating 3-brane resides in a higher dimensional
bulk with $n$ extra dimensions. If the bulk contains a positive cosmological constant, but no matter fields,
the brane worlds possess a curvature singularity at the origin. If matter fields
are included, periodic solutions which are regular at the origin are possible.
In the case of two extra dimensions, we have constructed solutions which 
 resemble the periodic vacuum solution outside the string core, but get modified due to the
presence of matter fields in the string core such that the solution is regular at the origin.
For three extra dimensions, the presence of the matter fields associated to the
local monopole in the bulk cannot regularise the solutions in a similar fashion.
However, we managed to construct periodic solutions which have additional
``branes'' appearing at some finite value of the bulk radial coordinate.

These scenarios are interesting since they provide models with a natural finiteness
of the extra dimensions in a by observations
confirmed natural setting of an inflating brane.

\ack Y.B. thanks the Belgian FNRS for financial support. B.H. was supported
by a CNRS grant. We thank the organisers
of the NEB XII conference for their hospitality.

\end{document}